\documentclass[fleqn,usenatbib]{mnras}

\usepackage{newtxtext,newtxmath}
\usepackage{color}

\usepackage[T1]{fontenc}

\DeclareRobustCommand{\VAN}[3]{#2}
\let\VANthebibliography\thebibliography
\def\thebibliography{\DeclareRobustCommand{\VAN}[3]{##3}\VANthebibliography}

\usepackage{graphicx}	
\usepackage{amsmath}	
\usepackage{tabularx}
\usepackage{listings}
\usepackage{anyfontsize}

\title[Constraining SHAM parameters]{Using a minimally parametrised SHAM to constrain the link between dark matter and galaxies}
\author[Fox L. Davidson et al.]{Fox L. Davidson,$^{1}$\thanks{E-mail: fox.davidson@port.ac.uk}
David Bacon, $^{1}$ 
Adam Amara,$^{2}$
Kazuya Koyama, $^{1}$ 
William G. Hartley, $^{3}$ \newauthor
L. F. de la Bella, $^{4}$
Sut-Ieng Tam, $^{5}$
Keiichi Umetsu, $^{6}$
Johannes Noller $^{7}$
\\
$^{1}$Institute of Cosmology and Gravitation, University of Portsmouth, Portsmouth, P01 3FX, UK.\\
$^{2}$School of Mathematics and Physics, University of Surrey, Guildford, Surrey, GU2 7XH \\
$^{3}$Department of Astronomy, University of Geneva, ch. d’Ecogia 16, CH-1290 Versoix, Switzerland\\
$^{4}$Faculty of Engineering and Physical Sciences, Surrey Space Centre, University of Surrey, Guildford, Surrey, GU2 7XH \\
$^{5}$Institute of Physics, National Yang Ming Chiao Tung University, No. 1001, Daxue Rd. East Dist., Hsinchu City 300093, Taiwan\\
$^{6}$Academia Sinica Institute of Astronomy and Astrophysics (ASIAA), No.1, Sec. 4, Roosevelt Rd, Taipei 106319, Taiwan\\
$^{7}$Department of Physics and Astronomy, University College London, Gower Street, London WC1E 6BT
}

\date{Accepted XXX. Received YYY; in original form ZZZ}

\pubyear{2024}

\begin{document}
\label{firstpage}
\pagerange{\pageref{firstpage}--\pageref{lastpage}}
\maketitle

\begin{abstract}
Models of the galaxy-halo connection are needed to understand both galaxy clusters and large scale structure. To make said models, we need a robust method that assigns galaxies to halos and matches the observed and simulated stellar-halo mass relation. We employ an empirical Subhalo Abundance Matching (SHAM) model implemented in the halos module of \texttt{SkyPy} which assigns blue and red galaxies based on the \citet{peng10} model containing three parameters: $M_\mu$ (halo mass where half the galaxies assigned should be quenched), $\sigma$ (transition width from star forming to quenched) and $b$ (baseline quenched fraction at low mass). We test two sets of galaxy stellar mass functions for four populations of galaxies (central/satellite, blue/red) and run parameter estimation using Approximate Bayesian Computation over each model when compared to a set of applicable literature models. For the \citet{weigel16} galaxies we find best fit values of log $M_\mu = 11.94^{+0.02}_{-0.02}$, $\sigma = 0.49^{+0.04}_{-0.04}$ and $b = 0.31^{+0.01}_{-0.01}$. For the \citet{birrer14} galaxies we find best fit values of log $M_\mu = 11.93^{+0.01}_{-0.01}$, $\sigma = 0.53^{+0.04}_{-0.04}$ and $b = 0.51^{+0.05}_{-0.04}$. Overall, we demonstrate that these constraints produce a model that is consistent with literature models for the central galaxies. Future research will focus on the normalisation of the satellite galaxies in order to better constrain the $b$ parameter.
\end{abstract}




\section{Introduction} \label{intro}




A complex part of modelling galaxy clusters and groups is to understand the link between dark matter and galaxies. Galaxies act as a biased tracer of the dark matter in the Universe as they formed when baryonic matter fell into the dark matter structure present when it decouples early in the Universe. However, the bias results in galaxies that do not trace the matter distribution exactly so we must understand the distributions statistically \citep{Desjacques_2018}. The galaxy-dark matter connection is important to understand as it unlocks insight into how both of these aspects of structure evolve in the Universe and can be explored in several different schemes of varying complexity. N-body simulations produce realistic structure formation for the dark matter \citep{klypin11, navarro97} but require hydrodynamical simulations of the gas and galaxies to be able to fully simulate a galaxy cluster and hence understand the link between dark matter and galaxies \citep{behroozi19, moster18, oleary20}. Both of these methods are computationally expensive.

Instead, computationally less expensive but more empirical models can be used to understand the galaxy-halo relation. These can be as limited as Halo Occupation Distribution (HOD), where only the number and distribution of galaxies are required \citep{kravtsov04, seljak00, skibba_sheth09}, to more complex Semi Analytic Models (SAMs) \citep{lacy_silk91, white_frenk91} which combine simplistic baryonic processes with N-body merger trees.

On the more empirical end lies Abundance Matching (AM). This is a method that makes the assumption of a monotonic relation between the mass or luminosity of a galaxy and its parent halo such that the most massive or luminous galaxy is assigned to the most massive halo. AM can be extended with the inclusion of subhalos (SHAM), plus the addition of the scatter inherent to baryonic processes and different assembly histories, to produce simulated galaxy clusters \citep{behroozi10, birrer14, moster10, stiskalek21}. Clusters are created by populating massive central halos with subhalos, by catalogues or N-body merger trees, and assigning galaxies as described above. Despite its lack of simulated physical processes, it has been shown that this method can still replicate the stellar mass - halo mass (SM-HM) relation \citep{wechsler18}.

The base SHAM technique will not differentiate between populations of galaxies. Real populations can be split primarily into two types: blue star forming galaxies and red quenched galaxies which have undergone some process that effectively stops their star formation, each modelled by a stellar mass function (SMF) \citep{schechter76}. A simplistic SHAM approach can be modified to assign galaxies of different colours or consider the fraction of red galaxies as a function of mass or other proxy property such as circular velocity or environment \citep{dragomir18, rodriguez_puebla15, yamamoto15}. \citet{peng10} \citep[extended in][to central and satellite galaxies]{peng12} explored the differential effects of mass and environment quenching i.e. whether a galaxy has been quenched by feedback processes dependent on the mass of the galaxy or environmental processes like ram pressure stripping in the satellite of a halo. These papers found that only satellite galaxies are affected by environment quenching but both satellite and central galaxies experience mass quenching. However the continuity equations explored in \citet{peng10} cannot adequately explain blue satellites so the model was furthered in \citet{de_la_bella21}. \citet{de_la_bella21} finds the same results as \citet{peng10} but expands the model to full differential equations for the number density of each observed population. The previous listed papers found that the fraction of red centrals or satellites follows a functional form similar to an error function, where the centrals only depended on stellar mass and the satellites have a dependence on both stellar mass and environment.

The aim of this paper is to apply the empirical model from \citet{peng10} and \citet{peng12} through abundance matching (linking a parent halo's mass with an appropriate galaxy's stellar mass) to produce realistic SM-HM relations. By matching previous SHAM paper results and using the assumptions of AM, we can therefore constrain the parameters for the quenching mass function for a given set of galaxy SMFs.

In this paper, we run \texttt{SkyPy} \citep{skypy21} to generate the catalogues of galaxies and halos and then apply the quenching model from \citet{peng10} and formalism of \citet{de_la_bella21} in a SHAM framework. The code presented here is available through the halos module of \texttt{SkyPy}. The mass quenching is controlled by two parameters: $M_\mu$ and $\sigma$, where $M_\mu$ is the characteristic mass where the quenching function predicts that half the galaxies will be quenched (i.e. the mean of the error function) and $\sigma$ is the standard deviation of the distribution, determining the width of the transition between the assigned blue and red galaxies around the mean halo mass. Changing these parameters allows us to parametrise the link between halo mass and the evolution of the assigned galaxies inside them. Environment quenching depends on the baseline $b$ which sets the minimum fraction of subhalos assigned quenched galaxies at low mass. This is due to subhalos having some probability of being quenched when they enter a parent halo.

We assume a flat $\Lambda$CDM model throughout with a Hubble parameter $h = 0.7$ ($H_0 = 70$ km s$^{-1}$ Mpc$^{-1}$), $\Omega_m = 0.3$, $\Omega_b = 0.045$, $\sigma_8 = 0.8$ and $n_s = 0.96$. All logs are log$_{10}$ and exp$(x)$ refers to the exponential function $e^x$. Halos may refer to any form of dark matter, parent halos refer to the dominant dark matter component in the system and subhalos are any further sub-components that may be present. Centrals and satellites refer to the galaxies residing in each of these halos. Blue, star-forming or unquenched galaxies refer to those galaxies that are still actively making stars; red or quenched galaxies refer to those that have stopped star formation. We include a clarification of what this means for both the galaxies we use and any models we compare to in Appendix \ref{Appred}. Finally, we refer to the collection of $M_\mu$, $\sigma$ and $b$ as the quenching parameters.

The structure of this paper is as follows. Section \ref{simpop} describes the galaxy and halo populations and models used throughout this work. Section \ref{sham} briefly describes the SHAM mechanism used (further described in Appendix \ref{AppSHAM}) as well as how the \citet{peng10} model is applied. In section \ref{results} we compare our model to previous literature and evaluate the goodness of our model as well as discuss our results about the constraining power on the quenching parameters. Section \ref{conclusion} contains our conclusions.



\section{Simulation populations} \label{simpop}
\subsection{Galaxy Populations} \label{galpop}
Different galaxy populations are modelled by stellar mass functions using a Schechter function \citep{schechter76}
\begin{equation} \label{equ:schechter}
    \text{d}n = \phi_* \text{exp} \left (-\frac{M}{M_*} \right) \left (\frac{M}{M_*} \right)^\alpha \text{d}M
\end{equation}
which gives the number density of a galaxy population in a given mass bin ($M$ to $M + \text{d}M$). Different populations of galaxies have different sets of parameters $M_*$ (characteristic mass), $\alpha$ (slope of the low mass power law) and $\phi_*$ (normalisation of the number density). Observationally, it has been shown that galaxies can be separated into four distinct populations: those that are star forming or quenched as stated before, but also those that are in the centre of a galaxy cluster (centrals) or those that are in the satellite of a cluster (satellites) \citep{weigel16, yang07}. For our main result in this paper we use the best fit Schechter functions given in table 6 of \citet{weigel16} for four populations: red centrals, red satellites, blue centrals and blue satellites. The fits for these galaxies are derived from observational results from SDSS where the spectral redshifts, spectra and magnitudes are used to estimate the stellar mass. The parameters for the Schechter function are then fitted, finding a single Schechter function for each. For additional results to show the potential optimisation for these Schechter functions we also use the $z=0$ parameters on tables 10 - 13 in \citet{birrer14} for the same populations. These were instead simulated as a gas regulator model and inserted into an N-body merger tree to find the SM-HM relation. Each galaxy population found from the previous process was fitted to the Schechter mass function to find the parameters given. We use the parameters derived for their Model C which is the closest they found to literature results. We describe each paper's method of dividing their galaxies in more detail in Appendix \ref{Appred}.

The four populations are generated using the sampling of their individual stellar mass functions through \texttt{SkyPy} in a narrow width of redshift  $0.01 \le z \le 0.1$ to match the observational data from \citet{weigel16} and the $z=0$ model in \citet{birrer14}. This also renders any redshift evolution negligible.

\subsection{Halo Populations}  \label{halpop}
In general, halo populations are modelled from theoretical halo mass functions (HMF) fitted to data or simulations, the earliest being the Press-Schechter function \citep{press_schechter74}. In later years these have been refined to more accurately reflect the populations observed in simulations \citep[such as][]{tinker08}. This paper uses the \citet{sheth_torman99} mass function to match the HMF used in \citet{birrer14}. We run \texttt{SkyPy} to directly sample our halo catalogue in a narrow width of redshift matching the galaxy sample.

\subsubsection{Subhalo generation} \label{subpop}
Subhalo populations in the late Universe are highly dependent on their host halo. Therefore we use the conditional mass function (CMF) derived in \citet{vale_ostriker04} (a modified Schechter function) to sample our subhalos
\begin{equation} \label{equ:V_O_CMF}
    N(m|M)\text{d}m = A \left (\frac{m}{x \beta M} \right )^{-\alpha} \text{exp} \left (-\frac{m}{x \beta M} \right ) \frac{\text{d}m}{x \beta M}
\end{equation}
where $m$ and $M$ are the subhalo and parent halo masses, $x$ is the factor by which the present day subhalos have been stripped of mass (i.e. if $x = 1$, $m$ is their present day mass whereas if $x > 1$, $m$ represents their initial mass before they became subhalos), $\beta$ is the exponential cutoff of the function as it approaches the parent's mass. $A$ is a normalising factor 
\begin{equation} \label{equ:V_O_norm}
    A = \frac{\gamma}{\beta \Gamma(2 - \alpha)}
\end{equation}
such that the sum of the subhalos masses is some fraction $\gamma$ of the parent's mass i.e. $\int_0^\infty m N(m|M) \text{d}m = x \gamma M$. It has been shown that the galaxy-halo relation correlates most to the maximum or pre-stripped mass of the subhalos \citep{wechsler18} so we therefore follow the example of \citet{vale_ostriker04} and use $x = 3$ when mass matching our subhalo population with their assigned galaxies. For the other parameters, we use $\alpha = 1.91$ and $\beta = 0.39$ \citep{vale_ostriker04} and $\gamma = 0.1$ as other studies have found that the total mass fraction of subhalos is closer to 10\% than the 18\% found in \citet{vale_ostriker04} \citep{gao11, de_lucia04}.

We use the occupation number (the number of subhalos a parent halo can support) to create our simulated satellite population
\begin{equation} \label{equ:V_O_occno}
    N_s = \frac{\gamma}{\beta \Gamma(2 - \alpha)} \Gamma(1 - \alpha, m_{\text{min}}/(\beta M))
\end{equation}
where $\Gamma(1 - \alpha, m_\text{min}/(\beta M))$ is the incomplete upper Gamma function and $m_\text{min}$ is the minimum mass for a subhalo to host a galaxy, hence, as mentioned in \citet{vale_ostriker04}, this generates the number of subhalos that can host galaxies. We set $m_{\text{min}} = 10^{10} M_\odot$ as this is the resolution of parent halos which we allow to generate subhalos.

\section{SHAM Mechanism} \label{sham}
The code used to generate the results in this paper is available through the halos module in \href{https://github.com/skypyproject/skypy}{\texttt{SkyPy}}. We include a brief summary of the model for reference. The SHAM model is run in the following way:
\begin{enumerate}
    \item  We run the \texttt{SkyPy} pipeline to generate the parent halo catalogue by sampling the chosen mass function through a YAML file which includes the cosmology and observational parameters
    \item The parent halos are used to generate the subhalo catalogue using equation \ref{equ:V_O_CMF} and \ref{equ:V_O_occno} in \citet{vale_ostriker04}. ID values that identify which subhalos are generated from which parent halos are created in this step
    \item The \citet{peng10} model, called the quenching function (equation \ref{equ:quench}), is applied to the parents and subhalos denoting them as hosting a quenched or unquenched galaxy
    \item We use the input SMF parameters to integrate each galaxy population's Schechter function (equation \ref{equ:schechter}) to some minimum mass such that the number of galaxies in 
    a population will approximately match the number of appropriately labelled halos/subhalos. We run the \texttt{SkyPy} pipeline to generate each population's catalogue using the SMF parameters, the previously found minimum mass and the same cosmology and observational parameters as the halos
    \item Optionally, we add scatter to the galaxies via a proxy mass, defined as a Gaussian scatter from the original galaxy. In this case the galaxies are mass ranked by the generated proxy mass and then assigned. Details for the scattering process are given in section \ref{scatter}
    \item Catalogues are mass ranked and matched. For the subhalos, we use their pre-stripped masses. Only appropriate galaxies are assigned to the dark matter, i.e. a blue central would be matched to an unquenched parent halo. Any unmatched halos or galaxies are ignored and not included in any further analysis
    \item The model outputs a dictionary containing the collective SHAM (arrays of galaxy, halo and stripped subhalo masses) and any identifiers (ID values, galaxy type etc.)
\end{enumerate}
Using the output dictionary, we can find SM-HM ratio as shown in this paper but further analysis could be made using the ID values to find cluster relations.

 In this section we will explain how the \citet{peng10} and \citet{de_la_bella21} model is applied but other details are explained further in Appendix \ref{AppSHAM}. The model is parametrised as a modified error function. The standard error function takes the form
 \begin{equation}
     \text{erf } z = \frac{2}{\sqrt{\pi}} \int_0^z \text{exp}(-t^2) dt
\end{equation}
where $z$ is some complex number and the function varies between -1 and 1. Our quenching function acts as the fraction of assigned quenched galaxies for a given halo mass $M$. We write the quenching function as
 \begin{equation} \label{equ:quench}
    q(M_\mu, \sigma, b, M) = \frac{(1-b)}{2} [1 + \text{erf}(r_0(M_\mu, \sigma, M))] + b \\
\end{equation}
where $M_\mu$, $\sigma$ and $b$ are the quenching parameters as previously described. erf is the error function with the input
\begin{equation} \label{equ:quench_inside}
    r_0 = \frac{\text{log} \left (\frac{M}{M_\mu} \right)}{\sqrt{2} \sigma}
\end{equation}
which transforms the error function to be centred at $M_\mu$ with a standard deviation of $\sigma$. We use the input variable $r_0$ to emphasise that the input is real and positive.

\begin{figure}
    \includegraphics[width=\columnwidth]{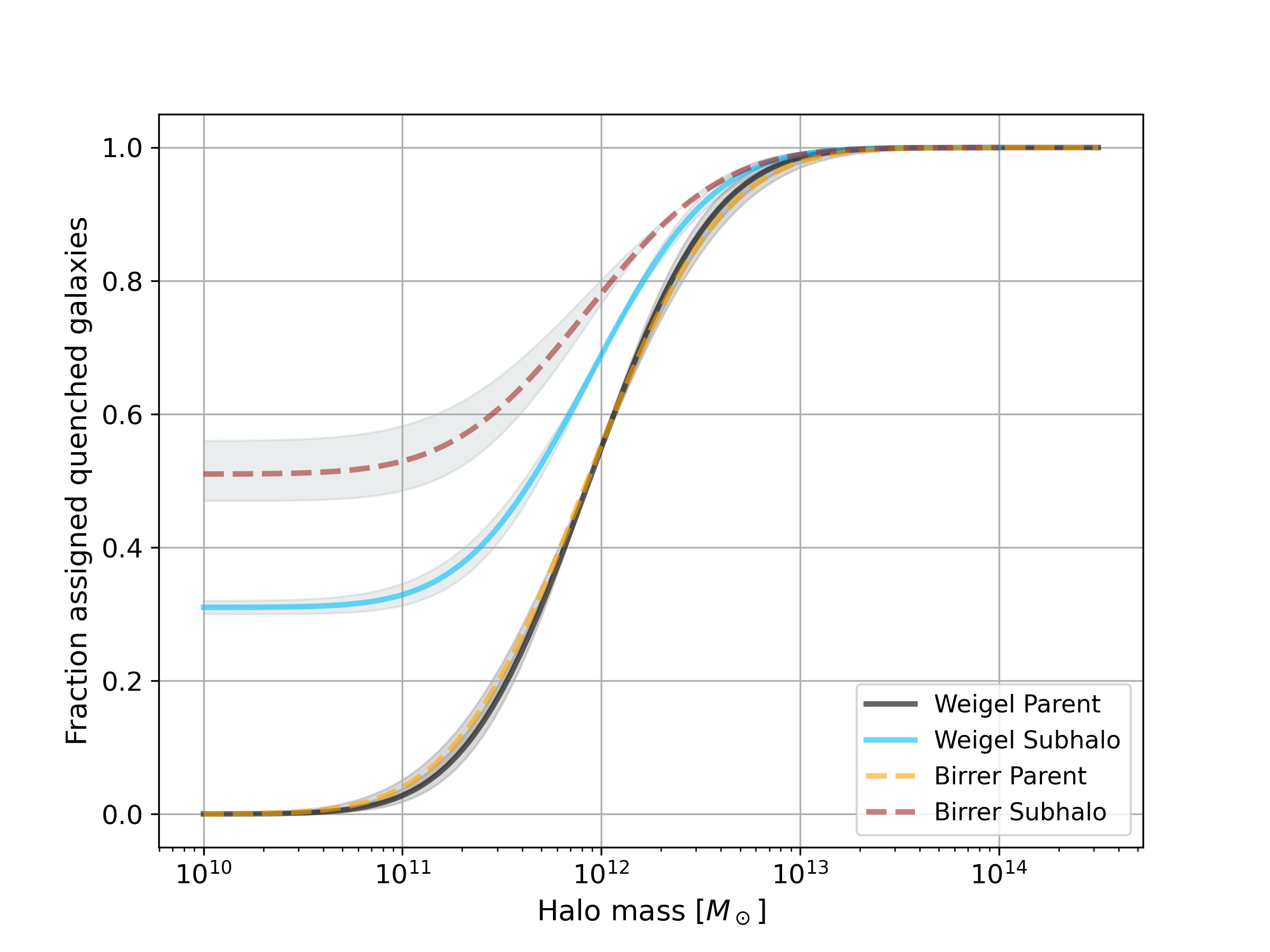}
    \caption{Quenching functions for both the parent and subhalos showing the best fit values of $M_\mu$, $\sigma$ and $b$ from table \ref{tab:best_fitwnbn}. The shaded areas are the 1$\sigma$ contours for each model as shown in the posteriors}
    \label{fig:quench_funct}
\end{figure}

 Equation \ref{equ:quench} ranges from 0 at low masses so that all halos below a given mass are assigned unquenched galaxies, to 1 at high masses so that only quenched galaxies are present above a given mass. For subhalos, we need to introduce a baseline i.e. the minimum fraction found at low mass. This is because subhalos must start as the parent halo of their own galaxy before becoming subdominant to another. This process adds the effect that every satellite has the potential to be quenched (the baseline fraction) when it falls into a larger halo. The subhalo function therefore varies from the baseline $b$ to 1 at high mass. A comparison of this function for the best fit values is seen in figure \ref{fig:quench_funct}.

\section{Results}  \label{results}
This section details how we compare our code written in \texttt{SkyPy} as a consistent model to literature and how we constrain the quenching function parameters.

\subsection{Comparison to literature} \label{comp}
To show that our framework gives realistic SM-HM relations, we have compared it to several results from the literature which range in technique. 
To constrain the quenching parameters we have compared our model to the models listed in table \ref{tab:models_list} where we list: location of the model values (for example the plot or equation it was generated from), location of the error in the models, the specific part of our model we compare to and any other considerations needed. The subhalo results use the maximum or accretion mass of the subhalos hence we use our pre-stripped subhalo masses for the relevant SM-HM relation. The redshift of the literature was also considered. Since we are modelling at very close to $z=0$ we have used models that fits this range and any redshift evolution in the literature parameters has been ignored where relevant.


As reference, because our model contains inherent randomness from both the sampling of distributions and the quenching function (although our averaged model's error is negligible compared to the error from the literature), we consider a `comparable model' to be three runs of the SHAM code with the same parameters and averaged in the literature value bins i.e. we find the average SM-HM ratio for the same halos given in each literature model.

We also include the consideration of needing a statistically significant number of model values in a given literature bin to be reasonably comparable. The plots in this paper show only the model values from the literature that were used in comparison with the model rather than the full set as provided in the papers. The majority of values removed are from the high mass end where the SM-HM relation is constant for the parameter space tested. This is because we generate very few high mass halos (as per the HMF) and hence defining a mean for those bins would not be statistically reasonable. A few low/high mass values are removed from the red/blue literature models respectively \citep{birrer14, rodriguez_puebla15} due to the low number of specific colour galaxies generated in the transition region near the peak of the SM-HM relation.

\begin{table*} 
	\centering
	\caption{Literature models list containing location of the model values, or equation and parameters for generation, location of the error values, what it is comparable to in our model and other considerations. Here maximum, accretion or infall mass is the wording taken from the given paper but they are treated in the same way in this paper.}
 \label{tab:models_list}
	\begin{tabularx}{0.824\textwidth} {|p{2.5cm}|p{2.5cm}|p{2.5cm}|p{2.5cm}|p{2.5cm}|}
		\hline
		  \textbf{Paper} & \textbf{Location of model values} & \textbf{Location of errors} & \textbf{Comparable} & \textbf{Considerations} \\
		\hline
		  \citet{moster10} & Equation 2, parameters table 1/3 & Fig 1, $1 \sigma$ shaded area, extracted & Centrals/Satellites & Satellites use maximum mass \\
		  \hline
            \citet{moster13} & Equation 2 \citep{moster10}, parameters table 1 Centrals/Satellites & Fig 12, $1 \sigma$ shaded area, extracted & Central/Satellites & Satellites use infall mass \\
            \hline
            \citet{behroozi10} & Equation 21, parameters table 2 & Fig 5, error bars, extracted & Centrals & - \\
            \hline
            \citet{behroozi13} & Equation 3, parameters section 5  & Fig 8, error bars, extracted & Centrals/Satellites & Satellites use accretion mass \\
            \hline
            \citet{rodriguez_puebla15} & Equation 17 \citep{behroozi13}, parameters equation 35/36 & Fig 5, scatter, extracted & Blue/Red Centrals & - \\
            \hline
            \citet{birrer14} & Fig 17, extracted (Model C) & Table 14, scatter on Model A (Model C not given) & Blue/Red/Average Centrals & - \\
            \hline
            \citet{rodriguez_puebla12} & Fig 2, extracted & Fig 1, generic $1 \sigma$ value, extracted & Satellites & Uses accretion mass \\
            \hline
	\end{tabularx}
\end{table*}
\subsection{Constraints on quenching parameters} \label{constraint}
We constrain the parameters by running a modified Bayesian Monte Carlo. The likelihood of the SM-HM is difficult to calculate so we use simulation based inference methods to avoid this calculation. We use the Approximate Bayesian Computation (ABC) method through the \texttt{abcpmc} \citep{abcpmc} Python package which is a Monte Carlo code that uses a threshold to progressively test values sampled from the prior and each iteration reduces the sample space to the `well-fitted' values (determined by the calculated $\chi^2$ of the averaged model values compared to each set of literature models) until it approaches the posterior. Our approach is similar to \citet{tam_2022} who used this method to find posteriors on cosmological parameters for synthetic weak lensing cluster catalogues.

\begin{figure}
	\includegraphics[width=\columnwidth]{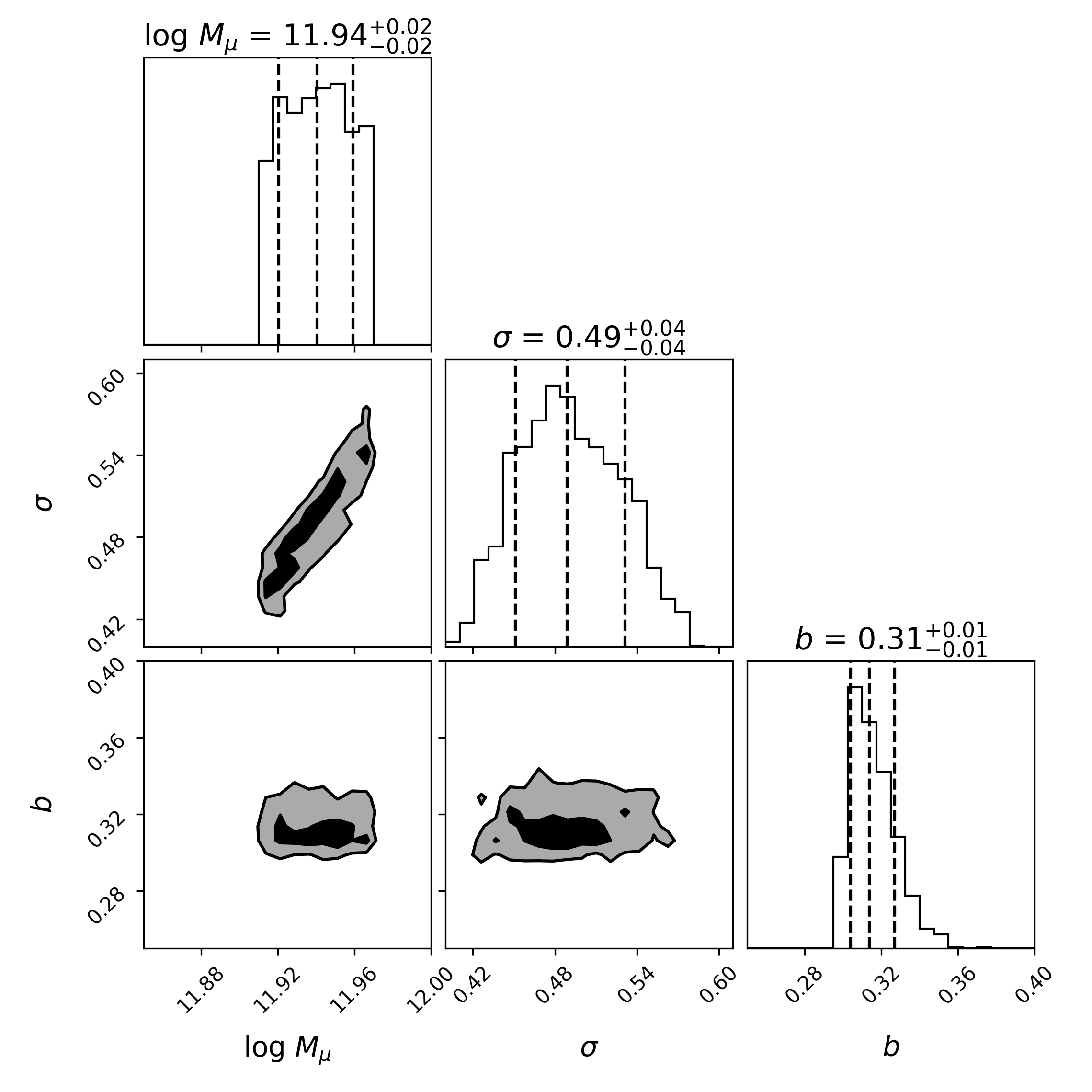}
    \caption{Posterior on parameter estimation for Weigel model without scatter. There is a significant degeneracy for the $\sigma$ parameter.}
    \label{fig:postwn}
\end{figure}

\begin{figure}
	\includegraphics[width=\columnwidth]{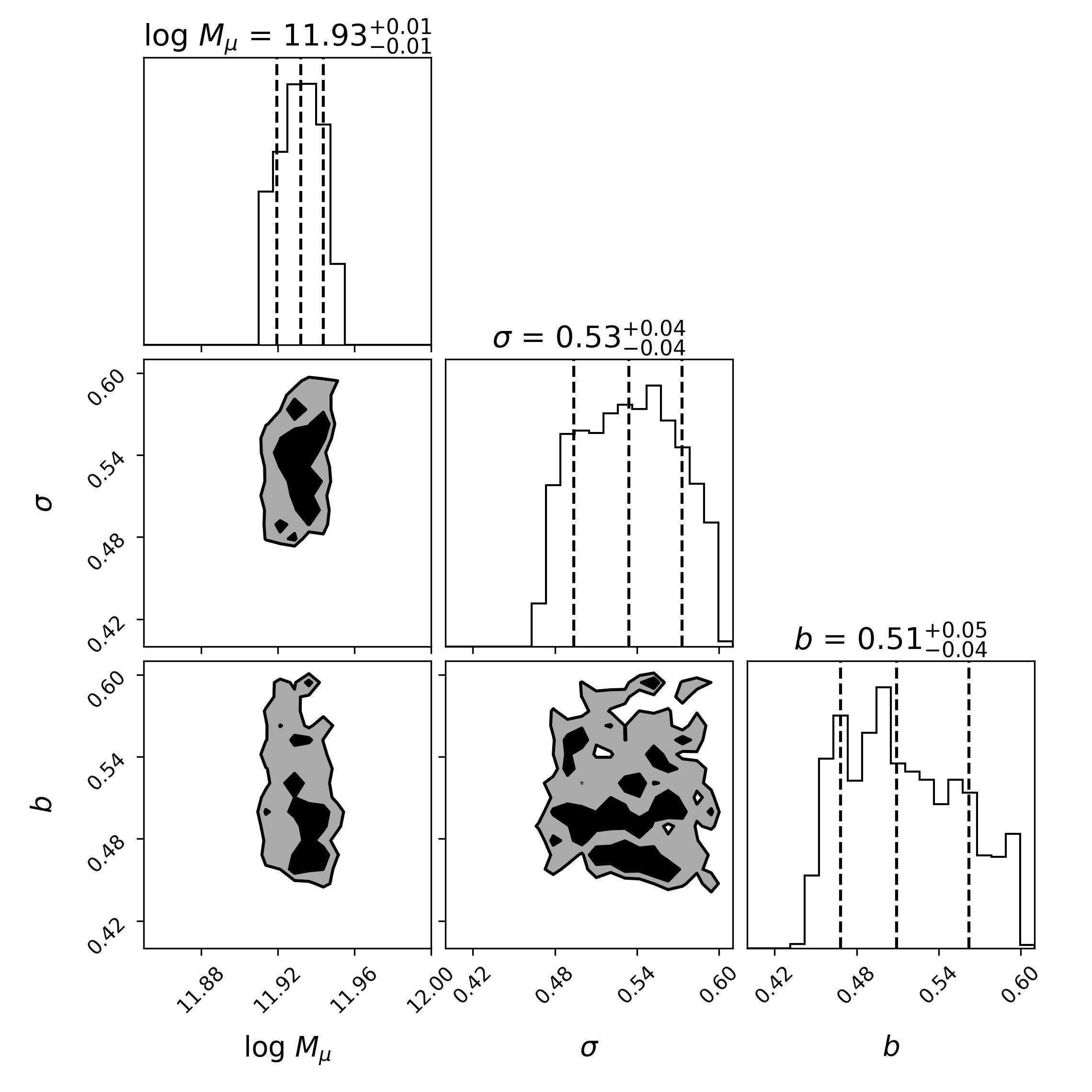}
    \caption{Posterior on parameter estimation for Birrer model without scatter. There is a significant degeneracy for the $\sigma$ and $b$ parameter.}
    \label{fig:postbn}
\end{figure}

\begin{figure}
	\includegraphics[width=\columnwidth]{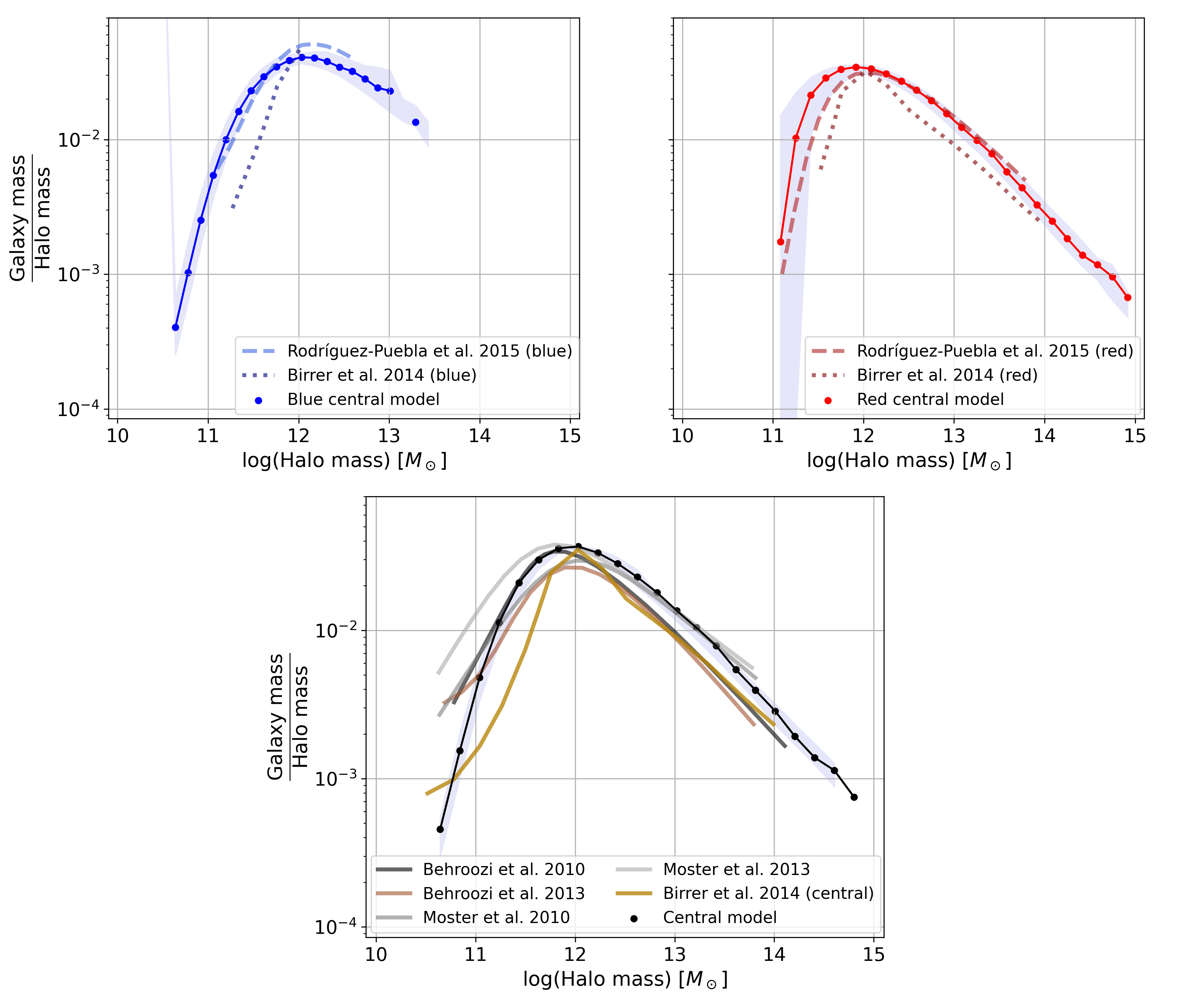}
    \caption{SM-HM relation for the centrals in the Weigel model. Connected scatter points indicate the model specified by the parameters in table \ref{tab:best_fitwnbn}. The top two plots show the blue and red centrals respectively along with the appropriate comparable models. The bottom plot shows the general centrals relation (average between the blue and red populations) and relevant literature models. The shaded regions are the range of models selected from the posterior distribution.}
    \label{fig:SM_HMwn}
\end{figure}

\begin{figure}
	\includegraphics[width=\columnwidth]{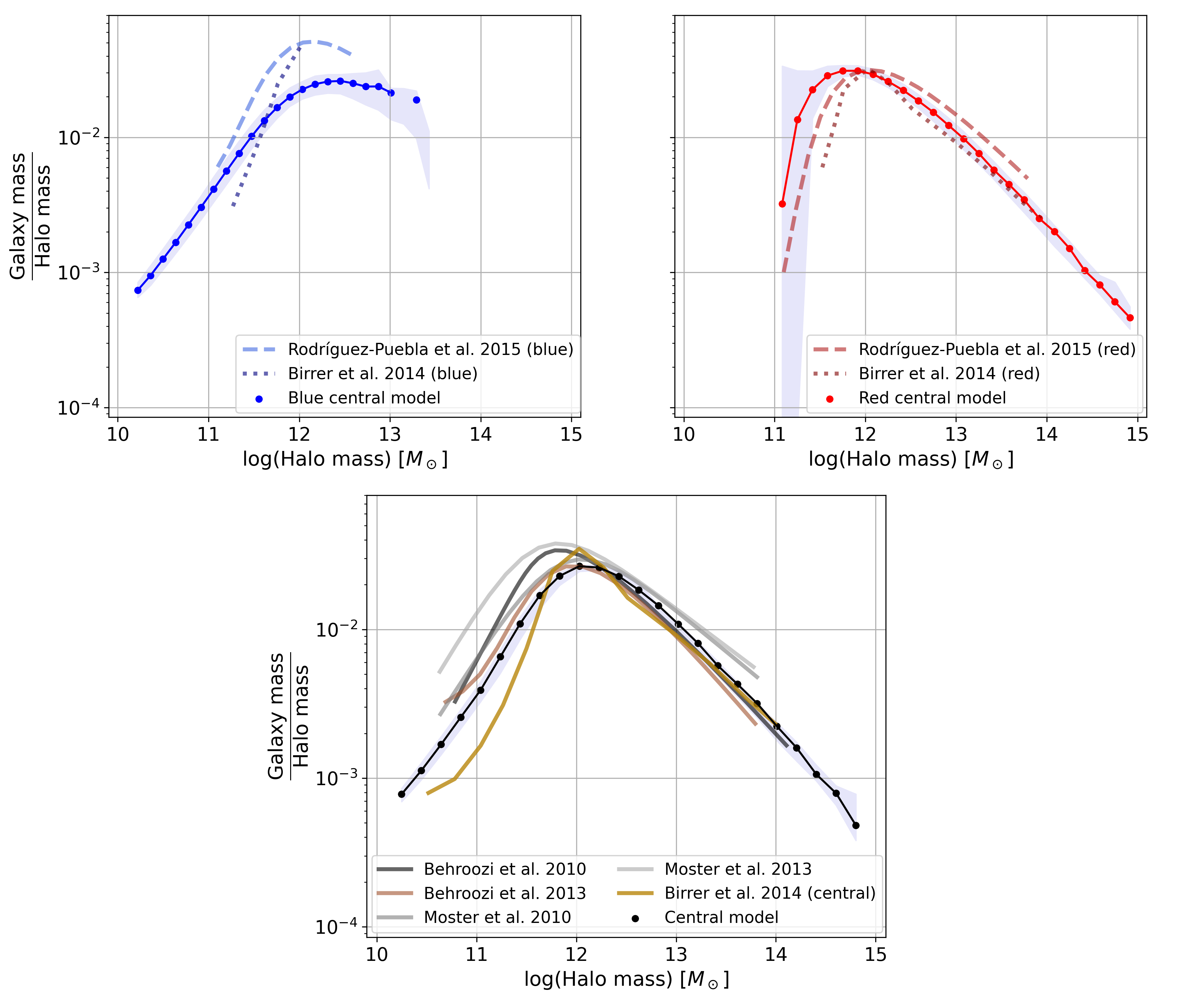}
    \caption{SM-HM relation for the centrals in the Birrer model. Connected scatter points indicate the model specified by the parameters in table \ref{tab:best_fitwnbn}. The top two plots show the blue and red centrals respectively along with the appropriate comparable models. The bottom plot shows the general centrals relation (average between the blue and red populations) and relevant literature models. The shaded regions are the range of models selected from the posterior distribution.}
    \label{fig:SM_HMbn}
\end{figure}

We ran 55 iterations against the models listed previously to find the best fit parameters in table \ref{tab:best_fitwnbn} which are shown in figures \ref{fig:postwn} and \ref{fig:postbn} for the Weigel and Birrer Schechter functions results respectively. We also show that these values generate models that fit the models well in figures \ref{fig:SM_HMwn} and \ref{fig:SM_HMbn}. We use strict priors on the values of log$(M_\mu)$ and $\sigma$ (11.91 - 11.97 and 0.4 - 0.58 respectively for the Weigel model, 11.91 - 11.95 and 0.47 - 0.6 respectively for the Birrer model) as these can be relatively well seen by eye but leave the prior for $b$ (0.3 - 0.5 for Weigel, 0.4 - 0.6 for Birrer) more open for reasons discussed in section \ref{change_smf}. We do this to increase the speed of the ABC iterations. We also use a prior to make sure that the red centrals are well-fitted since they are the most sensitive to the parameter space. We also include a prior for the satellites as certain $b$ values result in a discontinuity in the average satellite values which is clearly unphysical.


\begin{table}
	\centering
	\caption{Best fit values for the Weigel and Birrer models without scatter. We use the same value of $M_\mu$ and $\sigma$ for both parents and subhalos, but only the subhalos feel the effect of the base $b$ parameter.}
  \label{tab:best_fitwnbn}
	\begin{tabular}{llcr}
		\hline
		\textbf{Model} & log $M_{\mu}$ & $\sigma$ & $b$\\
		\hline
		  Weigel & 11.94$^{+0.02}_{-0.02}$ & 0.49$^{+0.01}_{-0.04}$ & 0.31$^{+0.01}_{-0.01}$ \\[0.1cm]
		  Birrer & 11.93$^{+0.01}_{-0.01}$ & 0.53$^{+0.04}_{-0.04}$ & 0.51$^{+0.05}_{-0.04}$ \\[0.1cm]
		\hline
	\end{tabular}
\end{table}

The posteriors for all tested galaxy SMFs show that our model is sensitive to the $M_\mu$ parameter but has degeneracies for the $\sigma$ parameter and the $b$ parameter for the Birrer model. Since $M_\mu$ controls the ``mid-point'' of the transition it has a stronger effect on which halos the quenched galaxies are assigned to, whereas for a given $M_\mu$ a reasonable number of $\sigma$s could bring the number of assigned quenched galaxies into line with the correct sampled number of galaxies. It is important to note that the averaged centrals (i.e. averaging the red and blue galaxies together) are relatively insensitive to both $M_\mu$ and $\sigma$ as only the central region near the peak in the SM-HM relation contributes to the parameter estimation. This is because the high mass red galaxies and the low mass blue galaxies will always be assigned to the correct halos following the shape of the quenching function (0 at low mass, 1 at high mass). It should also be noted that the red central population is the most strongly dependent on the quenching function as it is very sensitive to the change in abundance due to the shape of its Schechter function. While the blue centrals' SMF continues to increase at low mass and can hence generate as many galaxies as is necessary to match the assignment count, the red centrals' SMF decreases rapidly at low mass and hence cannot generate more than a set number of galaxies. This means that if the parameters generate a quenching function which produces too many or too few red assigned halos then the SMF cannot compensate and it produces a SM-HM relation of a significantly different shape than the literature.

The $b$ parameter also shows degeneracy but this is more due to a lack of well-fitting model. The normalisation of the SM-HM ratio is strongly dependent on the normalisation of the galaxy population and neither the Weigel or Birrer models do an adequate job of matching the literature in this respect. We talk about this more in section \ref{change_smf}.

\subsection{Adding scatter} \label{scatter}
\begin{figure}
    \includegraphics[width=\columnwidth]{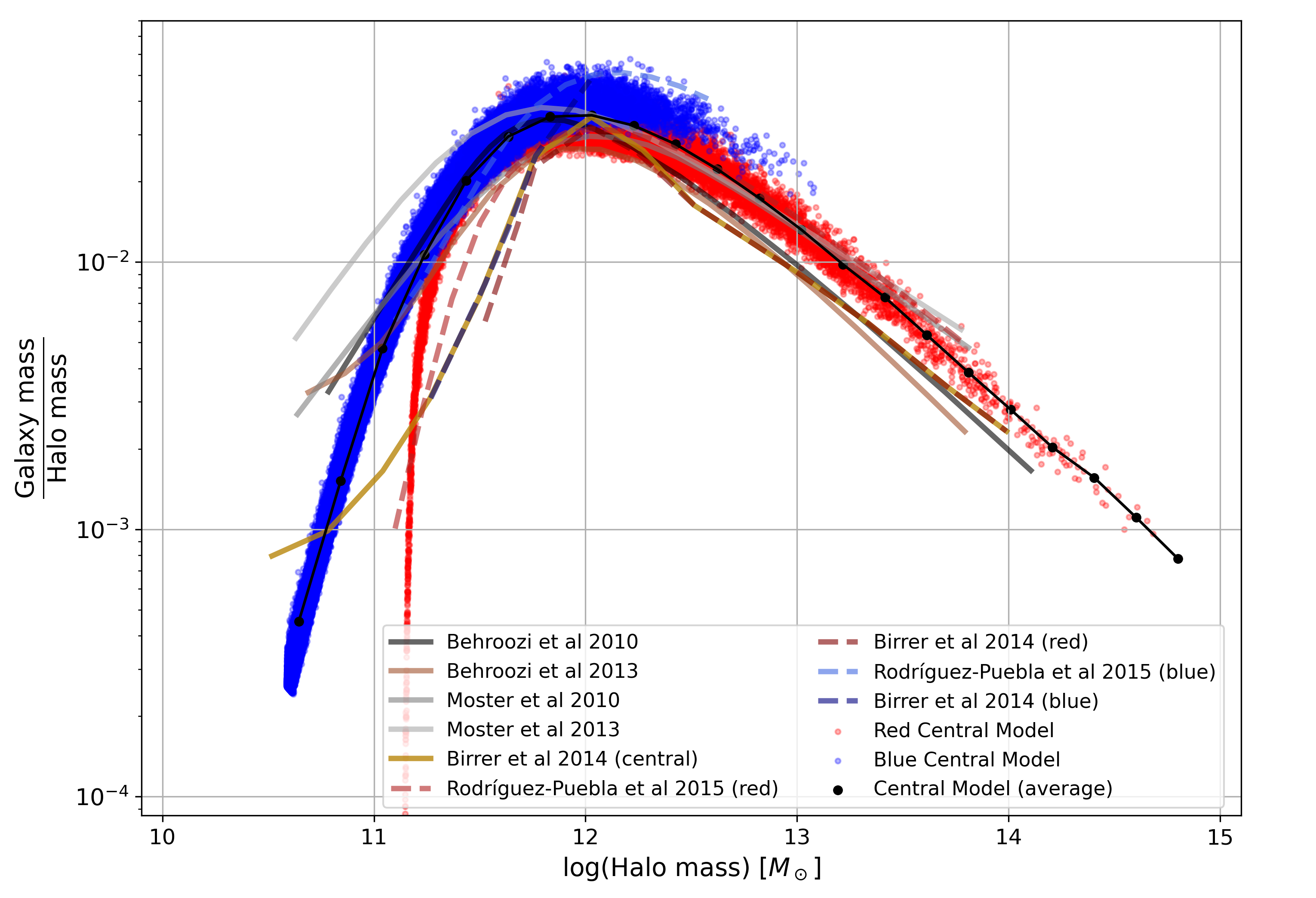}
    \caption{SM-HM relation for the central Weigel model with the parameters from table \ref{tab:best_fitwsbs}. Plotted are the sets of comparable models, the averaged centrals, all blue centrals and all red centrals to show the scatter in our models. The plot for the Birrer model is very similar.}
    \label{fig:SM-HM_scatter}
\end{figure}

\begin{figure}
    \includegraphics[width=\columnwidth]{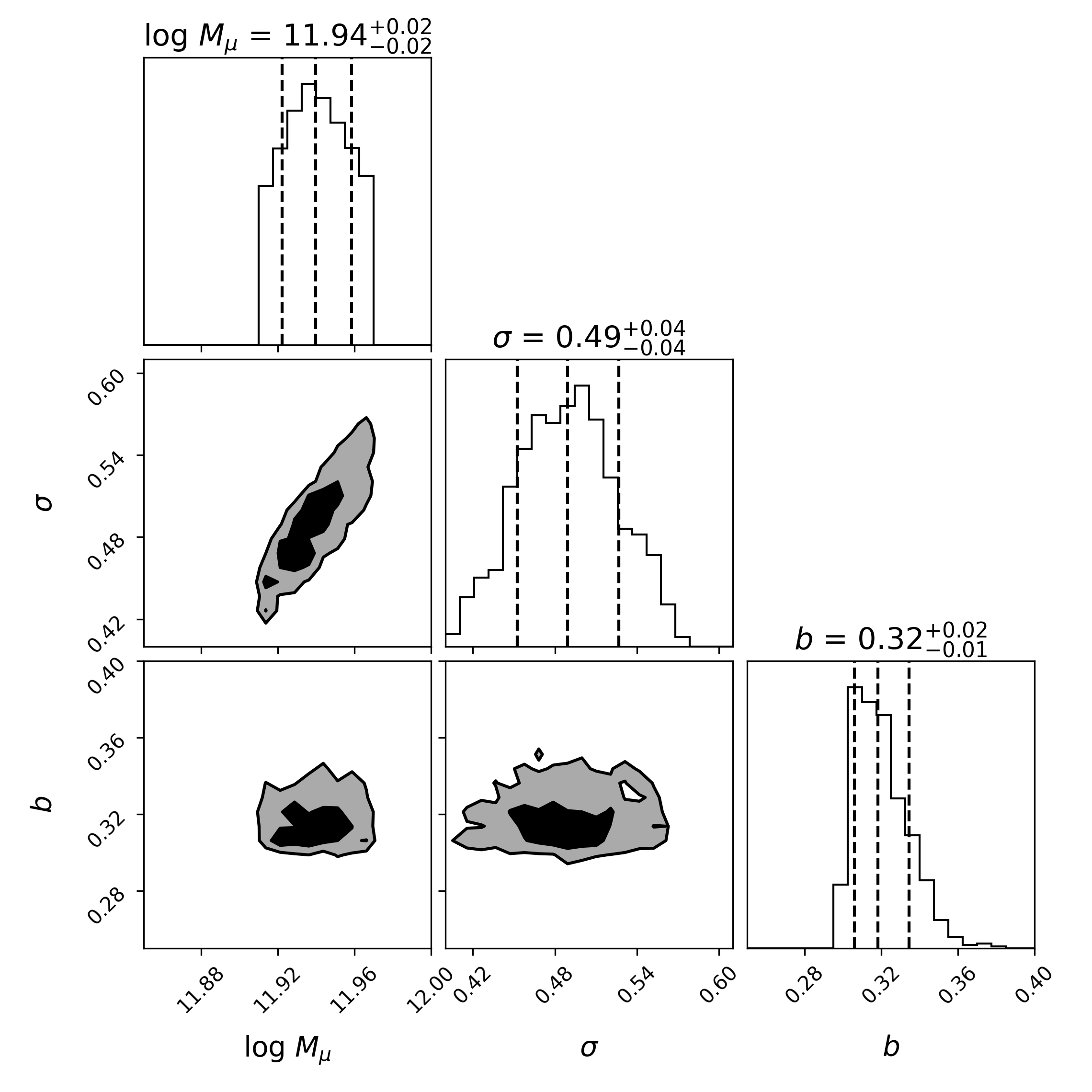}
    \caption{Posterior on parameter estimation for Weigel model with scatter.}
    \label{fig:postws}
\end{figure}

\begin{figure}
    \includegraphics[width=\columnwidth]{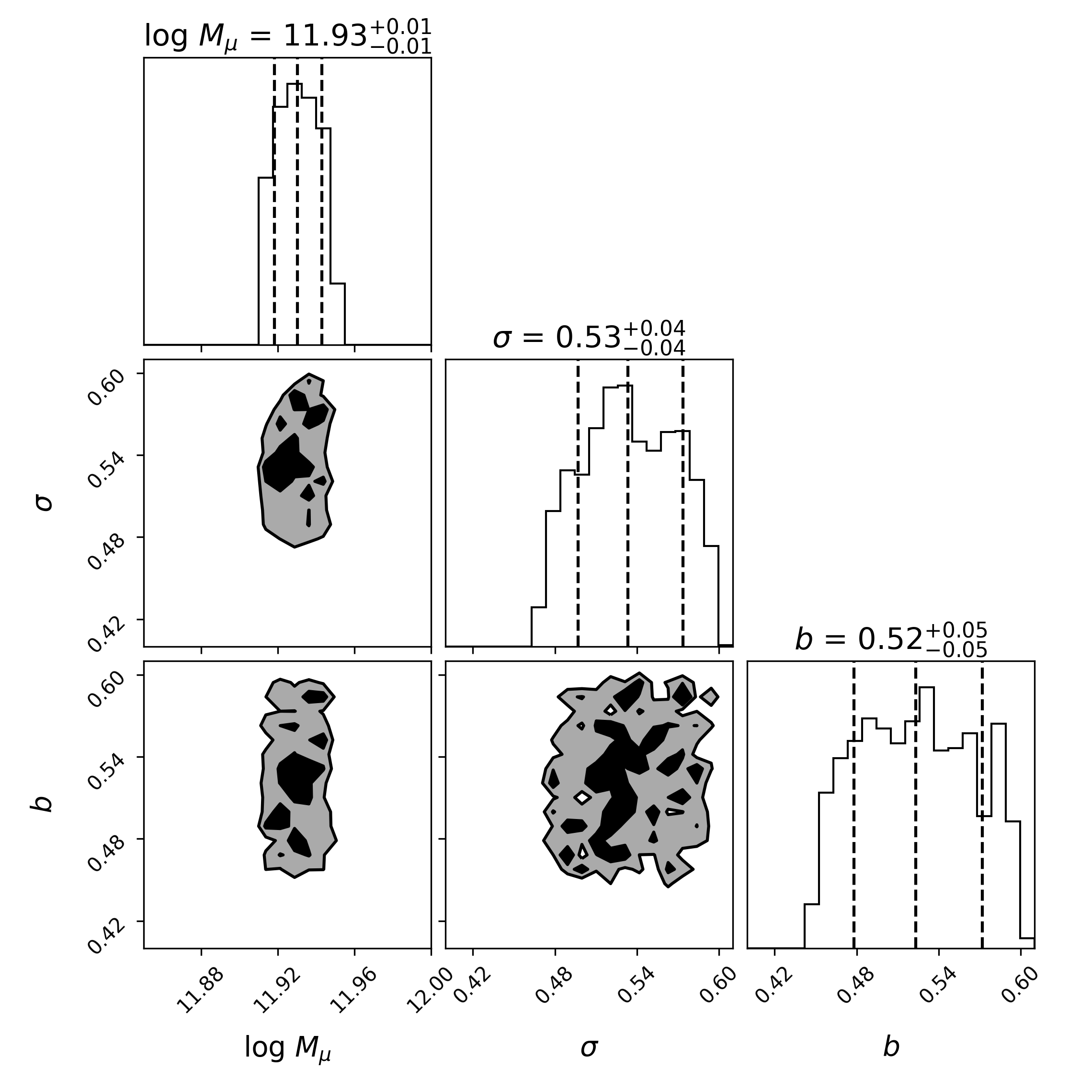}
    \caption{Posterior on parameter estimation for Birrer model with scatter.}
    \label{fig:postbs}
\end{figure}

A realistic model of the SM-HM relation includes scatter in the galaxy mass associated to a given halo mass. This is due to the different accretion histories that halos undergo in the Universe during structure formation. We have added scatter to our model via a proxy mass method. Once the galaxies are generated, we sample from a Gaussian using the galaxy's mass as the mean and 10\% of the mean as the standard deviation, so that it is constant in log space. This generates a uniform scatter in the SM-HM relation (figure \ref{fig:SM-HM_scatter}). The galaxies are then ordered by the proxy mass before being assigned. We run the fits again with the same procedure and find the parameters in table \ref{tab:best_fitwsbs} with posteriors in figures \ref{fig:postws} and \ref{fig:postbs}. Given the uniform scatter, we don't find a significant difference in the parameters as the averaged values that are compared to the models will be very similar.

\begin{table} 
	\centering
	\caption{Best fit values for the Weigel and Birrer models with scatter.}
 \label{tab:best_fitwsbs}
	\begin{tabular}{llcr}
		\hline
		\textbf{Model} & log $M_{\mu}$ & $\sigma$ & $b$\\
		\hline
		  Weigel & 11.94$^{+0.02}_{-0.02}$ & 0.49$^{+0.04}_{-0.04}$ & 0.32$^{+0.02}_{-0.01}$ \\[0.1cm]
		  Birrer & 11.93$^{+0.01}_{-0.01}$ & 0.53$^{+0.04}_{-0.04}$ & 0.52$^{+0.05}_{-0.05}$ \\[0.1cm]
		\hline
	\end{tabular}
\end{table}

\subsection{Tests on changing cosmology} \label{change_cosmos}
The literature models used to constrain the quenching parameters were generated with a mixture of different cosmologies either in a simulation or to calibrate masses. We test the robustness of our model by changing the fiducial cosmology to see if this affects the constrained quenching parameters' results.

In our model, the cosmology parameters will mostly affect the HMF and minimally affect the SMFs. Using the Planck 2018 `base $\Lambda$CDM' late Universe cosmological results \citep{planck18} ($h = 0.674$, $\Omega_m = 0.315$, $\Omega_b = 0.0493$, $ns = 0.965$, $\sigma_8 = 0.811$) we find that there is a negligible difference for the log$(M_\mu)$ parameter (0.024\% difference on average between the four models). However we find a larger difference for $\sigma$ (3.783\%) and $b$ (1.126\%). We still treat these as negligible because, as previously discussed, there is a significant degeneracy between the $M_\mu$ and $\sigma$ so a larger range of values for a well defined $M_\mu$ is expected. We should also expect a larger difference for $b$ because our model does not fit the literature well and hence the parameter is not well constrained, which we talk about in section \ref{change_smf}. For full clarity, the new results and the percentage difference for each model is shown in tables \ref{tab:planck} and \ref{tab:percent}. Given the minimal difference in the parameters with this cosmology to our main results, we expect that there would be equally negligible changes within the bounds of the \citet{planck18} cosmology. 

\begin{table} 
	\centering
	\caption{Best fit values for each model when using the \citet{planck18} cosmological results.}
 \label{tab:planck}
	\begin{tabular}{|l|c|c|c|}
		\hline
		  \textbf{Model} & log $M_{\mu}$ & $\sigma$ & $b$\\
		\hline
		  Weigel & 11.93$^{+0.02}_{-0.02}$ & 0.52$^{+0.03}_{-0.03}$ & 0.31$^{+0.01}_{-0.01}$ \\[0.1cm]
            Weigel + scatter & 11.94$^{+0.02}_{-0.02}$ & 0.52$^{+0.04}_{-0.04}$ & 0.31$^{+0.01}_{-0.01}$ \\[0.1cm]
            Birrer & 11.93$^{+0.01}_{-0.02}$ & 0.54$^{+0.04}_{-0.04}$ & 0.52$^{+0.05}_{-0.05}$ \\[0.1cm]
            Birrer + scatter & 11.93$^{+0.01}_{-0.01}$ & 0.54$^{+0.03}_{-0.04}$ & 0.51$^{+0.05}_{-0.05}$ \\[0.1cm]
		\hline
	\end{tabular}
\end{table}

\begin{table} 
	\centering
	\caption{Percentage difference on parameters for each model when using the \citet{planck18} cosmological results.}
 \label{tab:percent}
	\begin{tabular}{|l|c|c|c|}
		\hline
		\textbf{Model} & $\Delta$\% log $M_{\mu}$ & $\Delta$\% $\sigma$ & $\Delta$\% $b$ \\
		\hline
		  Weigel & 0.058 & 5.448 & 0.794 \\
		\hline
            Weigel + scatter & 0.027 & 5.843 & 1.253 \\
		\hline
            Birrer & 0.002 & 1.981 & 0.791 \\
		\hline
            Birrer + scatter & 0.011 & 1.858 & 1.667 \\
		\hline
	\end{tabular}
\end{table}

\subsection{Satellites} \label{change_smf}
\begin{figure}
    \includegraphics[width=\columnwidth]{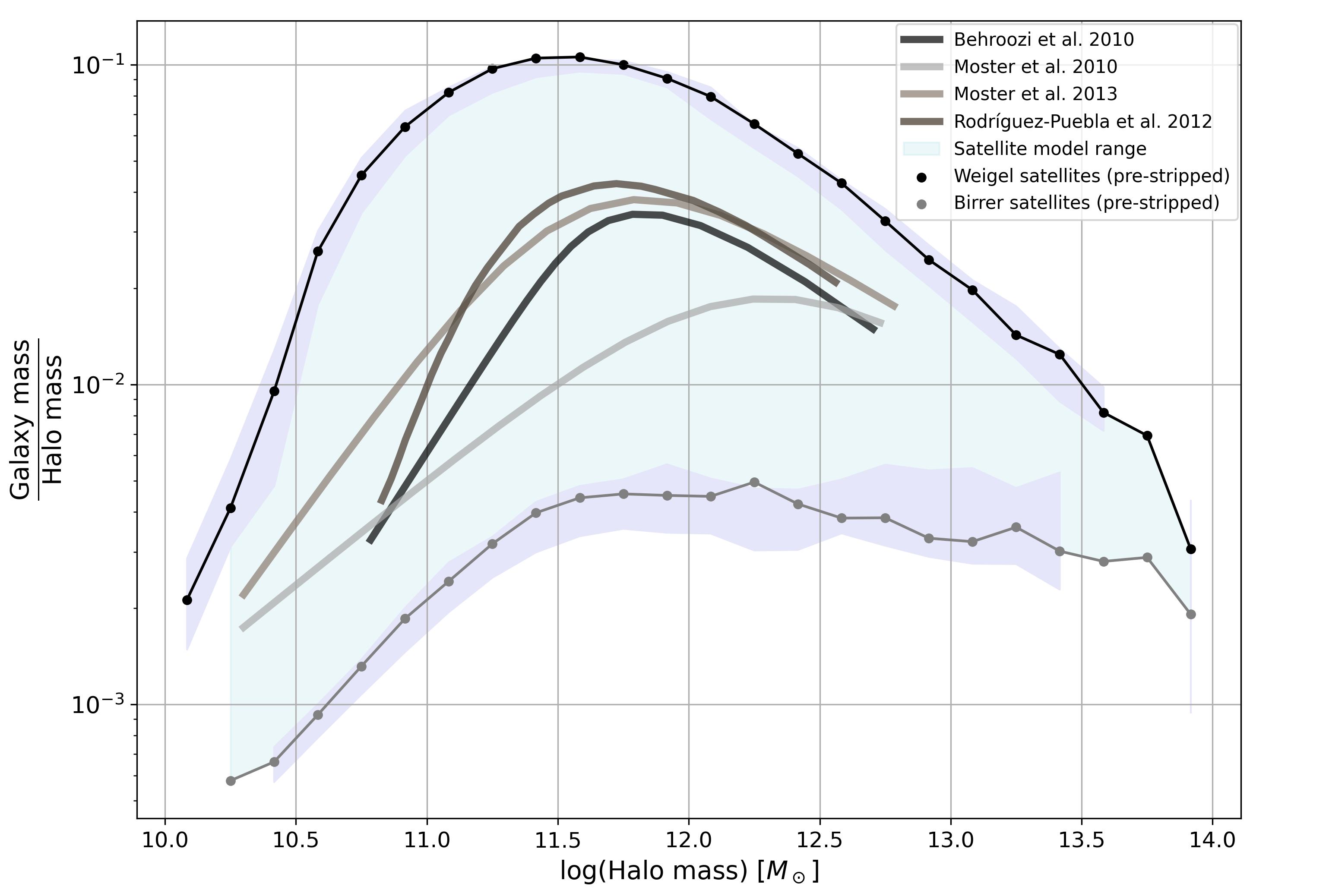}
    \caption{SM-HM relation for the both the Weigel and Birrer satellite models. Connected scatter points indicate the model specified by the parameters in table \ref{tab:best_fitwnbn}. The light blue shaded area shows the discrepancy between the normalisation of the two models and the literature and the purple shaded regions around each model are the range of models selected from their posterior distributions.}
    \label{fig:SM-HMsat}
\end{figure}

The model's SM-HM relation depends on relative mass distribution of the SMF and HMF as well as the normalisation of the SMF. Figure \ref{fig:SM-HMsat} shows the effect of using different SMFs (here from fitted observed galaxies in \citealt{weigel16} and simulated fitted galaxies in Model C of \citealt{birrer14}) on the satellite SM-HM relation. We can see that this changes both the shape and the normalisation of the SM-HM which is related to the normalisation of the SMF. Each model's SMFs have a much smaller effect on the centrals plot since they have similar SMFs for both red and blue centrals. This shows that the ABC method could be used to find the optimum SMF parameters for each population for literature models of satellite galaxies. While the $b$ parameter is well constrained (in the sense that the error values are small) this only indicates that these are values that do not introduce a discontinuity in the averaged satellite model and do not necessarily fit the models well given the distance between our models and the literature.

\subsection{Physical meaning of the quenching parameters} \label{meaning}
While the model we employ is very empirical, we can still consider the effect it has on the galaxy population and the implications on the physics. $M_\mu$ is the mean of the quenching function (equation \ref{equ:quench}) and hence defines the ``halfway'' point between the transition of blue and red galaxies. Because of this, it lies very close to the turn over in the SM-HM relation and hence close to the maximum star forming efficiency. $\sigma$ is the standard deviation and hence tells us about the width of transition between halos with only star forming and quenched populations. A small value of $\sigma$ would suggest that there is a sudden change around $M_\mu$ where the population becomes quenched, perhaps suggesting a threshold which leads to feedback or other quenching methods. A larger $\sigma$ suggests a slower transition where galaxies start to quench over a large halo mass range. Our results suggest that a mid range is preferred such that there is a significant number of purely blue or red galaxies on either side of the transition but the transition itself is not sharp. The $b$ parameter is the base fraction of satellite galaxies that are quenched as they enter a larger system. Larger values would suggest that the parent halo has properties such that it quickly strips the satellites, hence quenching them; smaller values suggest that there would not be much of an effect by entering another system. Since we apply this globally (i.e. not as a function of the parent halo mass each subhalo sits in) we can only comment on the averaged effect of a halo entering a larger system. While averaging between our models gives a value that means approximately half the low mass satellites should be quenched \citep[approximately matching the conclusions of][]{peng12}, the satellite galaxies used in our models have a different normalisation to the literature and hence we can only that these values create a physically plausible model i.e. they do not introduce a discontinuity in the SM-HM relation.

\section{Conclusions} \label{conclusion}
In this paper we have successfully constrained a minimally parametrised SHAM in order to model the galaxy-halo connection needed for large scale structure such as galaxy clusters. Our model allows us to form the SM-HM relation from a given HMF (parents) and generated subhalos, plus the SMF of four populations of galaxies (red/blue, central/satellite) to create galaxy clusters and groups. The different populations of galaxies are assigned based on a parametrised \citet{peng10} model by the quenching function (equation \ref{equ:quench}) which dictates the fraction of red galaxies for a given halo mass. This function depends on three parameters: $M_\mu$, the mean of the function or the halo mass at which half the assigned galaxies are red; $\sigma$, the standard deviation which controls the transition width between blue and red galaxy dominance; and $b$ the baseline quenched fraction of satellite galaxies. Mass quenching is felt by both central and satellite galaxies but environment quenching (controlled by the $b$ parameter) is only felt by the satellites.

By comparing our model to previous literature results we find that the quenching function appears to be closely related to the assumptions of abundance matching. We have constrained the parameters by using the ABC method to give best fit values for two different SMFs (table \ref{tab:best_fitwnbn}) that create a realistic SM-HM relation compared to previous literature for the central relations. We show that using different SMFs changes the normalisation of the SM-HM and the constrained parameters slightly due to the number of galaxies generated.

We also constrained the parameters for models with Gaussian scatter added to the galaxies (table \ref{tab:best_fitwsbs}). Due to the scatter being constant in log space, we find that the constrained parameters are similar to the previous results as the averaged model used when comparing to the literature is likely very similar to the non-scattered version. Testing our models with the \citet{planck18} cosmology showed a negligible change in the parameters given their constrainability.

Finally, we also considered the satellite model as this constrains the $b$ parameter. The two sets of SMF show the most difference in normalisation here but neither is well suited to the literature. Further work using this model could use the ABC method to find suitable SMFs to fit the literature's normalisation.

In summary, our results for the quenching function parameters are:
\begin{itemize}
    \item We find tight relations for the $M_\mu$ parameter in all cases, placing it close to the turnover in the SM-HM relation 
    \item Our constraint on the $\sigma$ parameter contains a degeneracy with $M_\mu$ and is hence less constrained. It is still well constrained for a given $M_\mu$ and gives an $\sim \mathcal{O}(2)$ of transition between blue and red centrals
    \item The $b$ parameter is not as well constrained due to the different normalisation between our two SMFs and the literature. We can say that the best fit is $\sim$ 0.5 which agrees with both \citet{peng10} and \citet{peng12} which is the basis for our quenching model. We can also conclude that it produces a physical model due to the lack of discontinuities in the SM-HM relation
\end{itemize}

These constraints show that we have produced a physical model which can reproduce literature results. The code for the model is available in the halos module 
 of \texttt{SkyPy} and further information is listed in the Appendix \ref{AppSHAM}.

\section*{Acknowledgements}
We acknowledge the funding of an STFC studentship in order to complete this research. We would like to acknowledge the help of our colleges in the SkyPy Collaboration, specifically I. Harrison and P. Sudek. We would also like to acknowledge the valuable conversation with Simon Birrer on the models presented in the \citet{birrer14} paper and conversations with Michael Kovac and Chayan Mondal. We also acknowledge the contributions of W. Enzi for important discussions about the \texttt{ABC} module code. K.U. acknowledges support from the National Science and Technology Council of Taiwan (grant NSTC 112-2112-M-001-027-MY3) and the Academia Sinica Investigator Award (grant AS-IA-112-M04)

We made the plots and code shown in this paper using the following Python packages: \texttt{NumPy}, \texttt{Visual Studio Code}, \texttt{SciPy} \citep{SciPy}, \texttt{Astropy} \citep{astropy}, \texttt{Matplotlib} \citep{matplotlib}, \texttt{Ipython/Jupyter} \citep{jupyter} and \texttt{corner} \citep{corner}. We were able to extract model values from the literature plots by using \texttt{WebPlotDigitizer} \citep{WebPlotDigitizer}. Numerical computations were done on the Sciama High Performance Compute (HPC) cluster which is supported by the ICG, SEPNet and the University of Portsmouth.


\section*{Data Availability}
The models used in this paper are taken from the literature and is available in the papers referenced. The specific literature values used to constrain this paper's model and featured in the plots in this paper are \href{https://github.com/Fox-Davidson/skypy/blob/halos_dev/skypy/halos/Literature_model_values%20(SHAM%20paper).txt}{listed with the code}. The model values generated in this paper are not available directly, but can be re-generated using the code available in \texttt{SkyPy}.




\bibliographystyle{mnras}
\bibliography{ms}




\appendix

\section{SHAM code} \label{AppSHAM}
The halos module of \texttt{SkyPy} can be accessed via the \href{https://github.com/Fox-Davidson/skypy/tree/halos_dev/skypy/halos}{Github page} and \href{https://skypy.readthedocs.io/en/latest/index.html}{documentation} is available. It can be installed by either cloning the git repository or using \texttt{pip} or \texttt{conda-forge} \\
\\
\verb|pip install skypy| \\
\\
\verb|conda install -c conda-forge skypy|\\
\\
When running the SHAM code, a \texttt{YAML} file detailing the functional form of the HMF \citep[see the \href{https://bdiemer.bitbucket.io/colossus/lss_mass_function.html}{\texttt{Colossus} documentation}][for further details]{colossus}, any cosmological and observational parameters (sky area and redshift). An example file would be
\begin{verbatim}
parameters:
  model: `sheth99'
  mdef: `fof'
  m_min: 1.e+9
  m_max: 1.e+15
  sky_area: 600. deg2
  sigma_8: 0.8
  ns: 0.96
cosmology: !astropy.cosmology.FlatLambdaCDM
  H0: 70
  Om0: 0.3
  name: `FlatLambdaCDM' #Requires user defined name
  Ob0: 0.045
  Tcmb0: 2.7
z_range: !numpy.linspace [0.01, 0.1, 100]
tables:
  halo:
    z, mass: !skypy.halos.colossus_mf
      redshift: $z_range
      model: $model
      mdef: $mdef
      m_min: $m_min
      m_max: $m_max
      sky_area: $sky_area
      sigma8: $sigma_8
      ns: $ns
\end{verbatim}

Here \texttt{z} and \texttt{mass} are the outputs of sampled redshifts and masses for the halos defined by the HMF, cosmology and volume. To run the SHAM code, the cosmology relevant to the galaxies must be specified and be the same as the halo \texttt{YAML} file as well as the halo file and galaxy parameters. As an example

\begin{lstlisting} [language=Python, showstringspaces=false]
import numpy as np
from astropy.cosmology import FlatLambdaCDM
from skypy.halos.sham import run_sham, sham_plots

h_file = `halo.yaml'
cosmology = FlatLambdaCDM(H0=70, Om0=0.3,
                          name='FlatLambdaCDM')
z_range = np.array([0.01, 0.1])
skyarea = 600

# Galaxy parameters
rc_p = [10**(10.75), 10**(-2.37), -0.18]
rs_p = [10**(10.72), 10**(-2.66), -0.71]
bc_p = [10**(10.59), 10**(-2.52), -1.15]
bs_p = [10**(10.59), 10**(-3.09), -1.31]
gal_p = [rc_p, rs_p, bc_p, bs_p]
qu = [10**(11.95), 0.48, 0.52]

sham_dict = run_sham(h_file, gal_p, cosmology,
                     z_range, skyarea, qu)

h = sham_dict['Halo mass']
g = sham_dict['Galaxy mass']
t = sham_dict['Galaxy type']
rc, rs, bc, bs, cen, sub = sham_plots(h, g, t)
\end{lstlisting}

These show the minimum number of parameters needed to run the code and there are many values (such as scatter, the CMF parameters etc.) that can be modified. \texttt{run\_sham} runs the SHAM code and outputs a dictionary containing the final information (see the documentation for more details) and \texttt{sham\_plots} uses it to output arrays of halo and galaxy masses for each population of galaxies to produce the plots seen in this paper.

\section{Red population} \label{Appred}
We include this appendix as clarification as to what a `red' or `quenched' galaxy means in the models we consider as well as for our galaxy catalogues. Our galaxy SMF parameters are from \citet{weigel16} and \citet{birrer14} which consider different ways to classify red galaxies. 

The galaxy sample obtained in \citet{weigel16} is from local galaxies surveyed in SDSS DR7. Galaxies are split into central and satellite by their environment, as provided by the \citet{yang07} sample. For colour, they use dust and K-corrected flux values from the NYU VAGC \citep{blanton05, padmanabhan08}. In the $u$ - $r$ stellar mass diagram, anything above a given $u$ - $r$ relation is considered to be red and anything below a similar relation is considered to be blue. Anything in the middle is denoted as the `green valley' which we do not include in this model.

In \citet{birrer14}, as mentioned previously they use a gas regulator system inserted into an N-body merger tree and they follow the evolution of the systems. The model itself does not separate between a quenched/unquenched galaxy; if the galaxy is unquenched, the gas regulator model continues, if it is quenched then it finishes and remains at that mass, producing no more stars. For mass quenching, a galaxy has some probability of becoming quenched when it gains some finite amount of mass and this probability is related to the change in mass. Environment quenching has a flat probability for any given satellite when it enters a larger system. This method does not change between the different models presented in the paper.

The other colour separated literature model is from \citet{rodriguez_puebla15}. They use a semi-empirical model to find the average central SM-HM from the SM-HM of the red/blue population and their fractions as a function of halo mass (their equation 19)

\begin{equation}
\begin{split}
    \langle \text{log} M_* (M_\text{h}) \rangle = 
    & f_b(M_\text{h}) \langle \text{log} M_{*,b} (M_\text{h}) \rangle + f_r(M_\text{h}) \langle \text{log} M_{*,r} (M_\text{h}) \rangle
    \end{split}
\end{equation}
where $\langle \text{log} M_{*} (M_\text{h}) \rangle$ is an average galaxy mass for a given halo mass  $M_\text{h}$ (our SM-HM relation multiplied by the halo mass) and $f(M_\text{h})$ is the fraction over the halo mass. $b$ or $r$ indicate the relations for blue and red centrals respectively. Their fractions use the formalism from \citet{peng12} but it is not parametrised in the same way as ours. Their SM-HM relations depend on the SMF of that particular population which they find by using local galaxies from the SDSS survey, specifically ones selected from \citet{yang12}. The colour is determined based on \citet{li06} colour-magnitude criteria using K-corrected colours.



\bsp	
\label{lastpage}
\end{document}